\documentclass[preprint]{elsarticle}

\usepackage{graphicx,subfigure}
\usepackage{epstopdf}
\usepackage{amsmath, amssymb}
\usepackage{rotating}
\usepackage{morefloats}

\begin{document}
\title{Spontaneous symmetry breaking of arbitrage}
\author{Jaehyung Choi\corref{cor1} }
\ead{jaehyung.choi@sunysb.edu}

\cortext[cor1]{Correspondence address: Department of Physics and Astronomy, SUNY at Stony Brook, NY 11794-3800, USA. Fax:+1-631-632-8176.}
\address{Department of Physics and Astronomy\\
 SUNY at Stony Brook, NY 11794-3800, USA}

\begin{abstract}
	We introduce the concept of spontaneous symmetry breaking to arbitrage modeling. In the model, the arbitrage strategy is considered as being in the symmetry breaking phase and the phase transition between arbitrage mode and no-arbitrage mode is triggered by a control parameter. We estimate the control parameter for momentum strategy with real historical data. The momentum strategy aided by symmetry breaking shows stronger performance and has a better risk measure than the naive momentum strategy in U.S. and South Korean markets.
\end{abstract}
\maketitle

\section{Introduction}
	After Bachelier's seminal paper \cite{Bachelier:1900p4113} and its re-discovery \cite{Cootner:1964p4779}, random walk theory has been the most crucial cornerstone in economics and finance. An assumption that price dynamics is governed by stochastic process has become popular and useful in asset valuation theories such as option pricing theory \cite{Black:1973p5045, Merton:1973p5006} or interest rate models \cite{Vasicek:1977p5439, Cox:1985p5352, White:1990p5282}. However, the assumption also claims that prices of financial instruments cannot be predicted exactly because of the nature of Brownian motion. This unpredictable nature of financial markets helps economists to establish a belief that there are no tools to find arbitrage opportunities and to make money systematically in the financial markets. It is also imposed that successful investors are considered nothing but luckier than others. The idea is crystallized in the form of the efficient market hypothesis by Eugene Fama \cite{Fama:1965p4897} and Paul Samuelson \cite{Samuelson:1965p4904}. According to the efficient market hypothesis, financial markets are informationally efficient and this efficiency cannot make participants systematically achieve excessive returns over the market portfolio in the long run. Although there are three slightly different versions of the hypothesis to cover more general cases, what the hypothesis generally emphasizes has not been changed.
		
	However, many market practitioners intrinsically have an idea that the market could be predictable regardless of their methods used for forecast and investment because it is partially or totally inefficient. The idea is opposite to the belief of proponents for the efficient market hypothesis and it is empirically supported by the fact that there are actual market anomalies which are used as the sources of systematic arbitrage trading. These anomalies and trading strategies include fundamental analysis, technical analysis, pair trading, price momentum, sector momentum, mutual fund arbitrage, volatility arbitrage, merger arbitrage, January effect, and weekend effect etc. The anomalies let market participants create profits by utilizing the trading strategies based on the market inefficiencies. Even if the market is efficient in the long run,  practitioners assure that they are able to find opportunities and timings that the market stays in the inefficient phase within very short time intervals. The existence of a shortly inefficient market state is guaranteed by the success of high frequency trading based on quantitative analysis and algorithmic execution in a short time scale automated by computers. In these cases, the arbitrage does not have the traditional definition that non-negative profit is gained almost surely. It can create positive expected return with high probability but there are also downside risks which make the portfolio underperform. This kind of arbitrage is called statistical arbitrage and the arbitrage in this paper means mostly statistical arbitrage.
	
	Not only the practitioners but some academic researchers also have different opinions to the efficient market hypothesis. They have taken two approaches to check the validity of the efficient market hypothesis. On the one hand, the market anomalies of which the practitioners believe the existence are empirically scrutinized. Some results on the famous market anomalies are reported in academic literatures and seem to be statistically significant while their origins are not clearly revealed yet. For more detailed discussions on the market anomalies, see Singal \cite{Singal:2006p5475} or Lo and MacKinlay \cite{Lo:2001p5538}. On the other hand, psychological and behavioral aspects of investors begin to be paid attention in order to find the explanatory theories on the market anomalies \cite{Shleifer:2000, Kahneman:2000, Kahneman:1982}. The behavioral economists focus on cognitive biases such as over- and under-reaction to news/events and bounded rationality of investors \cite{Conlisk:1996p1448}. They claim that those biases can create the inefficiencies in the market. The cognitive biases lead the investors to group thinking and herding behavior that most of investors think and behave in the similar ways. The good examples of herding are speculative bubbles, their collapses, market crashes, and panics during financial crises.
	
	Momentum effect on price or valuation in assets is one of the famous examples which have attracted the interest of industry and academia. As an implemented trading strategy, momentum strategy is frequently used for statistical arbitrage. Additionally, an investor gains a maximum 1.31\% of monthly return by the monthly momentum strategy that constructs the portfolio which buys past winners and short-sells losers in the U.S. market \cite{Jegadeesh:1993p200}. Since price dynamics has a tendency that price moves along the direction it has moved, price momentum becomes the systematic proxy for forecasting future prices. If an investor buys past winners, short-sells past losers, and repeats execution of the strategy, then he/she is expected to gain positive return with high probability in the long run. It is exactly a counterexample to the efficient market hypothesis. Despite the success of the momentum strategy, the origin of price momentum is not well-understood and remains rather unclear. Some possible explanations on the existence of the momentum effect answer parts of the question and the behavioral direction is one of them for understanding the nature of price momentum.
	
	Physicists also have become interested in the characteristics of financial markets as complex systems. Mainly, econophysics and statistical mechanics communities have used their methodologies to analyze the financial markets and several research fields have attracted their interests. In the sense of correlation, the financial markets are interesting objects. Since there are many types of interactions between market building blocks such as markets-markets, instruments-instruments, and investors-investors, correlations and correlation lengths are important. In other directions, speculation and its collapse are always hot topics because they are explained as collective behavior in physics. The analysis on speculation gives some partial answers that speculations have patterns including the resilience effect. Additionally, market crash or collapse of a bubble can be understood by the log-periodic pattern. For more details, see \cite{Mantegna, Roehner, Sornette,Sornette:2001p32,Malevergne:2001} and references therein.
	
	In particular, Sornette introduced the concept of spontaneous symmetry breaking (SSB) of stock price to explain speculation and to resolve the growth stock paradox \cite{Sornette:2000p4733}. He pointed out that economic speculation is understood as price dynamics caused by desirable/undesirable price symmetry. If stocks of a certain company are desirable to hold, investors try to buy the equities at extremely high prices which are the spontaneous symmetry breaking mode. However, when the equities are not desirable any more, the investors do not want to hold it and try to sell them as soon as possible to avoid damages from the downslide of price caused by the situation that nobody in the market prefers the equities. In his paper, the phase transition is induced by riskless interest rate above risk-adjusted dividend growth rate which also expresses herding in the sense that large growth rate gets more attention from investors and it leads to herding. Positive dividend payment breaking the symmetry makes the price positive and this is why the positive price is observed. These are the origins of speculation in economic valuation. The result is also related to the well-known financial valuation theory called the Gordon-Shaprio formula. His work is important in speculation modeling not only because symmetry breaking concept is applied to finance but also because speculation, its collapse, and market crash are indispensable parts of the market dynamics.
	
	In this paper, the concept of spontaneous symmetry breaking is applied to arbitrage modeling. Unlike Sornette's work \cite{Sornette:2000p4733} which uses spontaneous symmetry breaking to explain speculation in the asset valuation theory, the phase transition is emergent directly from arbitrage dynamics. Wyarta and Bouchaud also consider symmetry breaking \cite{Wyarta:2007} but their concern is self-referential behavior explained by spontaneous symmetry breaking of correlation in macroeconomic markets such as indexes not of arbitrage return generated by the trading strategy. From the viewpoint of symmetry breaking, this paper pays attention to portfolio/risk management rather than explanations on macroeconomic regime change on which both of the previous works focus. Based on the dynamics which gives a spontaneous arbitrage phase and a no-arbitrage phase, the arbitrage strategy can be executed upon the phases of arbitrage. The phases are decided by a control parameter which has the same meaning to speed of adjustment in finance. The execution of the strategy aided by spontaneous symmetry breaking provides better performance than the naive strategy and also diminishes risk of the strategy.  In Section \ref{SSB}, a brief introduction to arbitrage modeling is given and then the spontaneous arbitrage modes are emergent from the return dynamics. The momentum strategy aided by spontaneous symmetry breaking is simulated on real data and the results in various markets are posted in Section \ref{Result}. In Section \ref{Conclu}, we conclude the paper with some discussions and future directions.

\section{Spontaneous symmetry breaking of arbitrage}
\label{SSB}
\subsection{Arbitrage modeling}
	Introducing the existence of arbitrage opportunity, the value of portfolio $\Pi$ is governed by the following differential equation,
	\begin{eqnarray}
		d\Pi\big(t,r(t)\big)=\big(r_f+r(t)\big)\Pi\big(t,r(t)\big) dt+\sigma(t)\Pi\big(t,r(t)\big)dW(t) \nonumber
	\end{eqnarray}
	where $r_f$ is risk free rate, $r(t)$ is excessive return of the portfolio $\Pi$, $\sigma(t)$ is volatility of portfolio return, and $W(t)$ is a stochastic noise term. If the no-arbitrage theorem is imposed, the excessive return becomes zero guaranteed by the Girsanov theorem that the risk-neutral measure $\widetilde{\mathbb{P}}(t)$ and the Brownian motion $\widetilde{W}(t)$ always exist under no-arbitrage situation \cite{Girsanov}. If the existence of arbitrage is assumed, there is no risk-neutral measure  $\widetilde{\mathbb{P}}(t)$ nor related Brownian motion $\widetilde{W}(t)$. In this case, it is more important to know how its return series has evolved. The reason why the dynamics is important has two facets. First of all, for theorists, the dynamics encodes large amount of information on market macro- and microstructure. Secondly, it is helpful for practitioners to exploit the arbitrage opportunity by implementing trading strategies based on the dynamics. 
	
	The excessive return $r(t)$ is modeled by
	\begin{eqnarray}
		\label{arb_gen_dyn}
		\frac{dr(t)}{dt}=f\big(r(t)\big)+\nu(t)
	\end{eqnarray}
	where $\nu(t)$ is a white noise term. The structure of $f\big(r(t)\big)$ is decided by properties of arbitrage. One of the simplest forms for $f(r)$ is a polynomial function of $r$. Two properties of arbitrage dynamics help to guess the structure of the function \cite{Ilinski:2001p4046}. When the excessive return of the strategy is large enough, the arbitrage opportunity usually disappears very quickly because many market participants are easily able to perceive the existence of the arbitrage and can use the opportunity profitably even with trading costs. This property imposes a constraint that coefficients of $f(r)$ have negative values. Additionally, Eq. (\ref{arb_gen_dyn}) should be invariant under parity transformation $r\to-r$ because negative arbitrage return is also governed by the same dynamics. This property makes even order terms in the function vanish. Considering these properties of arbitrage, the form of $f(r)$ is given by
	\begin{eqnarray}
		\label{arb_dyn_poly}
		f(r)=-\lambda_1 r-\lambda_3 r^3-\cdots
	\end{eqnarray}
	where $\lambda_i>0$ for odd positive integer $i$. In traditional finance, these $\lambda$s are also able to be considered as the proxies incorporating the information on changes of discount rates which are covered in \cite{Cochrane:2011}. The dynamics describes reversal of return that the return becomes decreased when being large and it is increased when under the trend line. In other words, the reversal makes the return stay near the equilibrium around the trend line. By dimensional analysis, $\lambda_1$ is a speed of adjustment and is broadly studied in finance \cite{Amihud,Damodaran,Theobald:1998,Theobald:1999}. Larger $\lambda_i$ means the arbitrage opportunity dies out much faster. Meanwhile, smaller $\lambda_i$ corresponds to the situation that chances for arbitrage can survive longer. As $\lambda_i$ goes to infinity, the arbitrage return goes to zero extremely quickly and this limit corresponds to the no-arbitrage theorem. When only the linear term is considered for the simplest case, the dynamics is an Ornstein-Uhlenbeck process in mathematical finance,
	\begin{eqnarray}
		dr_t=(\mu-\lambda_1 r_t)dt+\sigma dW_t \nonumber
	\end{eqnarray}
	where the trend line $\mu$ is zero. This stochastic differential equation is  invariant under parity transformation of $r_t$ because $W_t$ is an Ito process with standard normal distribution which has symmetric distribution around mean zero. Although there are higher order terms in Eq. (\ref{arb_dyn_poly}), the dynamics is still considered as a generalized Ornstein-Uhlenbeck process because it is the mean-reverting process around the trend line.
	
\subsection{Asymptotic solutions}
	We begin to introduce a cubic term to the Ornstein-Uhlenbeck process to extend it to more general cases. The introduction of higher order terms is already used in the market crash model \cite{Bouchaud:1998p1217}. Then the dynamics is changed to
	\begin{eqnarray}
		\frac{dr(t)}{dt}=-\lambda_1 r(t)-\lambda_3 r^3(t)+\nu(t) \nonumber
	\end{eqnarray}
	where $\lambda_1>0$, $\lambda_3>0$, and $\nu(t)$ is a white noise term. After the cubic term is introduced, adjustment on arbitrate return occurs quicker because the coefficients are all negative. The negative coefficient condition needs to be modified in order to describe not only reversal but also trend-following arbitrage return which is explained by positive coefficients. In real situations, the trend-following arbitrage strategies are also possible to make profits by exploiting market anomalies because arbitrage opportunities fortified by transaction feedback do not disappear as quickly as expect and there could be more chances for investors. Speculation, as one of the examples, can create more opportunities for the trend-following arbitrage and increases expected return. Under speculation, the investors buy the instrument even though the price is high. This transaction induces to generate the trend line and is able to give feedback to the investors' trading patterns. During market crash or bubble collapse, they want to sell everything at very low prices although the intrinsic values of instruments are much higher than the price at which they want to sell. Not in extreme cases but under the normal market condition, people tend to buy financial instruments which have shown better performance than others because they expect that the instruments will provide higher returns in the future. The prices of the instruments become higher because the investors actually buy with the expectation \cite{Hong:1999p4506,Terence:1998p4385}. It seems to be very irrational but happens frequently in the markets. To integrate these kinds of situations, we can introduce the cutoff value which can decide whether the arbitrage is originated from reversal or trend-following dynamics rather than the negative speed of adjustment. With the cutoff value, let us change $\lambda_1$ and $\lambda_3$ into the forms of
	\begin{eqnarray}
		\lambda_1&\to&\lambda-\lambda_c \nonumber\\
		\lambda_3&\to&\lambda_c/r^2_c \nonumber
	\end{eqnarray}
	where $\lambda$, $\lambda_c$, and $r_c$ are positive. Although the number of parameters seems to be increased, this is not true because $\lambda_c$ is an external parameter. Under these changes, the arbitrage dynamics is given by
	\begin{eqnarray}
		\label{arb_dyn_ssb}
		\frac{dr(t)}{dt}=-(\lambda-\lambda_c)r(t)-\lambda_c\frac{r^3(t)}{r^2_c}+\nu(t).
	\end{eqnarray}
	
	After relaxation time $\tau$, Eq. (\ref{arb_dyn_ssb}) becomes zero up to the noise term because other transient effects die out. In other words, the deterministic part of arbitrage dynamics arrives at the equilibrium state. By setting the deterministic part of the r.h.s. in Eq. (\ref{arb_dyn_ssb}) to zero, stationary solutions are found. The interesting point is that the number of stationary solutions is dependent with $\lambda$ and $\lambda_c$. In the spontaneous symmetry breaking argument, $\lambda$ is a control parameter and $r$ is an order parameter. When $\lambda \ge \lambda_c$, there is only one asymptotic solution $r(t>\tau)=0$ which shows the property of usual arbitrage opportunities. The meaning of this solution is that the arbitrage return finally becomes zero up to noise. It is obvious that the arbitrage opportunity vanishes after the relaxation time because it is taken by market participants who know the existence and use the chance. 
	
	For $\lambda<\lambda_c$, there are three asymptotic solutions with $r(t>\tau)=0$ and
	\begin{eqnarray}
		r(t>\tau)=\pm\sqrt{1-\frac{\lambda}{\lambda_c}}r_c=\pm r_v. \nonumber
	\end{eqnarray}
	The solution $r=0$ has the same meaning to the solution for $\lambda \ge \lambda_c$. It means that the arbitrage opportunity finally dies out. The latter solutions, $r=\pm r_v$, are more interesting because there exist long-living arbitrage modes in return. After the relaxation time, the arbitrage chance still exists and lifetime of the spontaneous market anomaly is longer than that of the usual short-living arbitrage. It is noteworthy that these solutions unlike $r=0$ are symmetry breaking solutions although the dynamics is conserved under parity. The spontaneous mode also has the coherent meaning in the sense of speed of adjustment $\lambda$. If $\lambda$ is smaller than the critical value $\lambda_c$, it is slower adjustment and the arbitrage opportunity can have longer lifetime. These solutions are also well-matched to the no-arbitrage theorem that the arbitrage chance does not exist because it disappears very quickly. The no-arbitrage theorem which corresponds to $\lambda \to \infty$ does not make the arbitrage possible after the relaxation time because $\lambda$ is always greater than $\lambda_c$.
	
	When a weak field term is introduced to Eq. (\ref{arb_dyn_ssb}), the observation becomes more interesting. Introducing the constant term $\rho$, the equation is given by
	\begin{eqnarray}
		\frac{dr(t)}{dt}=\rho-(\lambda-\lambda_c)r(t)-\lambda_c\frac{r^3(t)}{r^2_c}+\nu(t) \nonumber
	\end{eqnarray}
	where $\rho$ can be considered the velocity of $r$. If $\lambda<\lambda_c$,  the asymptotic solution is also changed from $r_v$ to $-r_v$ as positive $\rho$ is changed to negative.
	
\subsection{Exact solutions}
	The asymptotic behaviors described in the previous subsection can be cross-checked with exact solutions. In the long run, the noise term is ignored because its average is zero. Under this property, the exact solutions of Eq. (\ref{arb_dyn_ssb}) are given by
	\begin{eqnarray}
		\label{ssb_exact}
		r(t)=\pm r_v\frac{r(t')\exp{(-(\lambda-\lambda_c)(t-t'))}}{\sqrt{r^2_v-r^2(t')(1-\exp{(-2(\lambda-\lambda_c)(t-t'))})}}
	\end{eqnarray}
	where $t'$ is the initial time. When $\lambda \ge \lambda_c$, exponential functions in the nominator and the denominator go to zero in the large $t$ region and it makes $r(t)$ zero. This corresponds to the symmetry preserving solution which is the usual arbitrage. If $\lambda<\lambda_c$, the exponential functions become dominant as $t$ goes to infinity. At that time, $r(t)$ approaches $\pm r_v$ which are the symmetry breaking solutions. These solutions are already seen in the asymptotic solutions.
	
\begin{figure}[!h!t!b]
		\subfigure[]{\includegraphics[width=6cm]{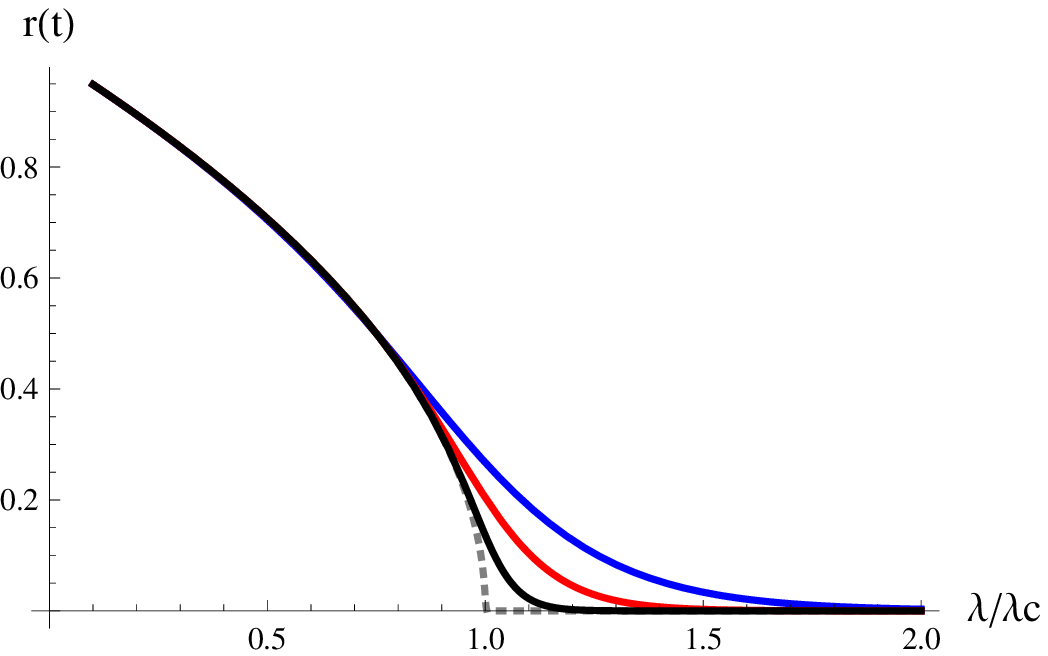}}
		\subfigure[]{\includegraphics[width=6cm]{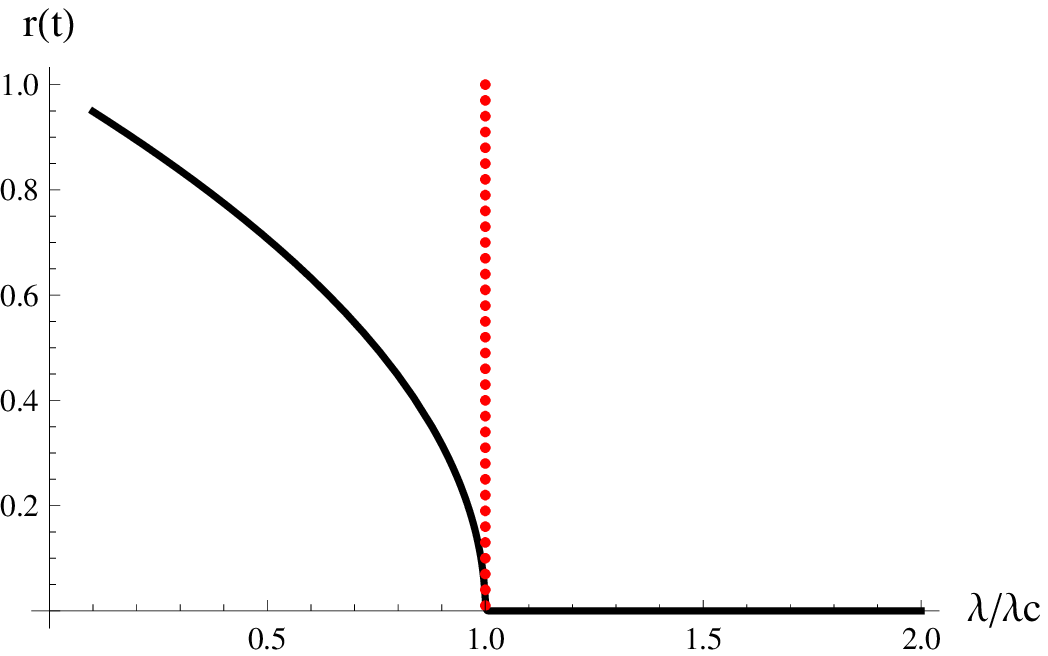}}
		\caption{Return vs. $\lambda/\lambda_c$. In the left graph, t=5 (blue), t=10 (red), t=25 (black), and t=$\infty$ (gray dashed). In the right graph, t=$\infty$ (black) and $\lambda/\lambda_c=1$ (red dotted)}
		\label{ssb_return_lambda}		
	\end{figure}

\begin{figure}[!h!t!b]
		\subfigure[r(0)$>r_v$]{\includegraphics[width=6cm]{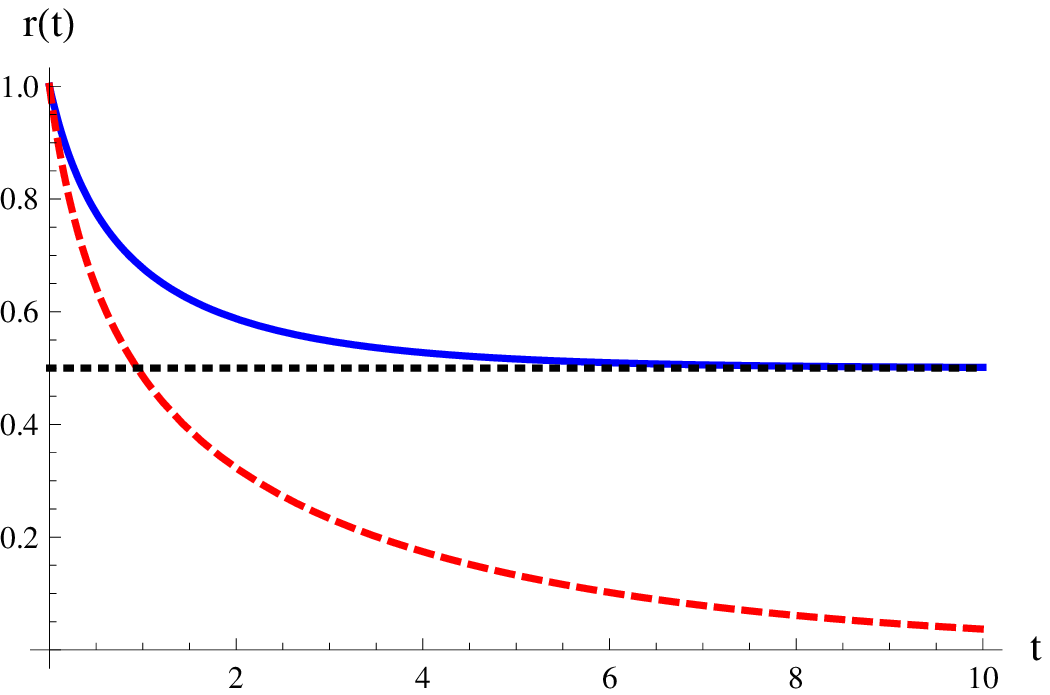}}
		\subfigure[r(0)$<r_v$]{\includegraphics[width=6cm]{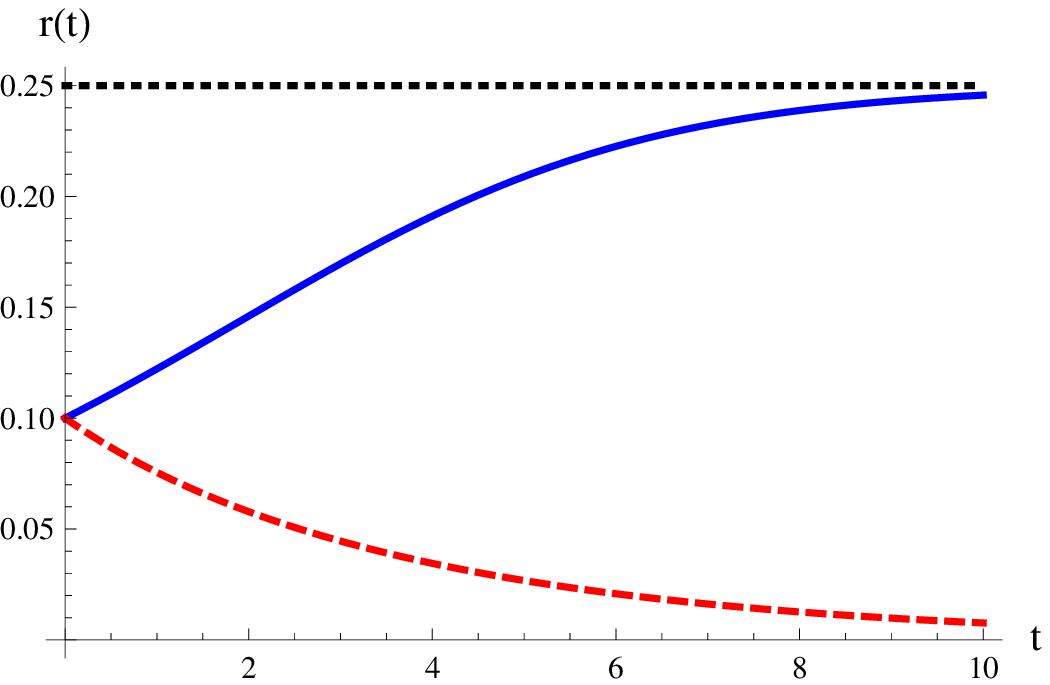}}
		\caption{Return vs. time. long-living arbitrage mode (blue), short-living arbitrage mode (red dashed), and asymptotic return (gray dashed)}
		\label{ssb_return_time}		
	\end{figure}

	With the long-living arbitrage solutions in Eq. ($\ref{ssb_exact}$), properties of the solutions are checked graphically in Fig. \ref{ssb_return_lambda} and \ref{ssb_return_time}.
	
	In Fig. \ref{ssb_return_lambda}, the left graph shows time evolution of the solutions as $t\to\infty$. In the small t region, there exist non-zero arbitrage returns regardless of the value of $\lambda/\lambda_c$. However, as $t\to\infty$, the return approaches to non-zero if $\lambda/\lambda_c < 1$ and it vanishes if $\lambda/\lambda_c \ge 1$. In the asymptotic region, the difference becomes clear and phase transition happens where $\lambda$ is at the critical value $\lambda_c$. It is easily seen in the graph on the right. The region $\lambda/\lambda_c < 1$ is called the long-living arbitrage phase, spontaneous return phase, or arbitrage phase. Another region where $\lambda/\lambda_c\ge1$ is considered the short-living arbitrage phase or no-arbitrage phase. In the model, market anomalies survive if they are in the long-living modes.
	
	In Fig. \ref{ssb_return_time}, the spontaneous arbitrage returns approach to $r_v$ whatever initial return values are. However, the no-arbitrage phase finally goes to zero. This property does not depend on the  size of the initial return values. Even if the initial value is smaller than the asymptotic value, it grows up to the asymptotic value. For example, if investors realize the arbitrage opportunity and if they begin to invest into the chance, their trading behavior affects price dynamics and the trend-following investors pay attention to the instruments. The interest leads to trading which gives feedback to their trading patterns and can increase the profitability. In other words, money flows into the instrument, boosts its price, and gives feedback to investors' behaviors. If transaction cost is smaller than the asymptotic value, arbitrage opportunities created by spontaneous symmetry breaking can be utilized by the investors.

	When the long-living arbitrage mode is possible, $r(t)$ can be re-parametrized by
	\begin{eqnarray}
		r(t)=\pm r_v+\psi_\pm(t) \nonumber
	\end{eqnarray}
	where $\psi(t)$ is a dynamic field for expansion around $\pm r_v$. Plugging this re-parametrization into (\ref{arb_dyn_ssb}), the differential equation for $\psi$ is solved and its solutions are given by
	\begin{eqnarray}
		\psi_\pm(t)=0, \mp\frac{2r_v}{1-\exp{(-(\lambda_c-\lambda)t)}} \nonumber
	\end{eqnarray}
	Since the latter solution goes to $\mp 2r_v$ in the asymptotic region, we can check the transition between $r_v$ and $-r_v$. If $\psi=0$, the initial modes stay in themselves, i.e. $\pm r_v$ go to $\pm r_v$. However, if $\psi$ is the latter solution, they evolve to $\mp r_v$ in large $t$ limit even though we start at $\pm r_v$ initially.
	
\section{Application to real trading strategy}
\label{Result}
\subsection{Method and estimation of parameters}
\begin{figure}[!h!t!b]
		\begin{center}
		\includegraphics[width=10cm]{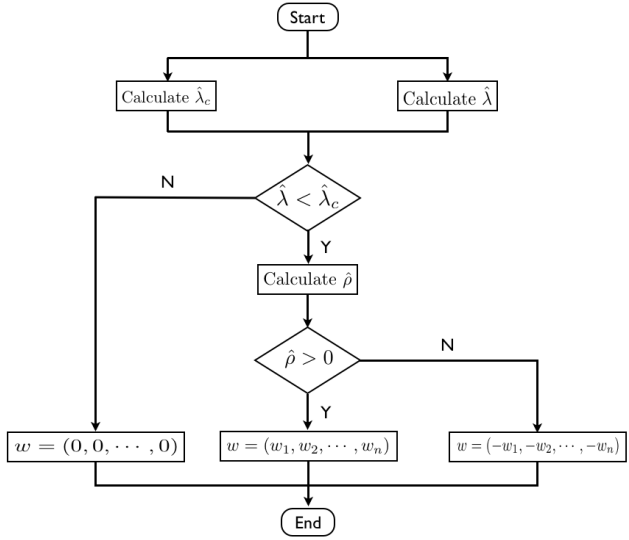}
		\end{center}
		\caption{Flow chart of the scheme based on spontaneous symmetry breaking concept}
		\label{flow_chart}		
	\end{figure}

	In order to test the validity of spontaneous symmetry breaking of arbitrage, we apply the following scheme depicted in Fig. \ref{flow_chart} to trading strategies over real historical data. In backtest, the control parameter $\lambda$ for the strategy should be forecasted based on historical data. At certain specific time $t'$, it is assumed that data only from $t<t'$ are available and the control parameter for next period is forecasted from them. If the forecasted $\lambda$ is smaller than the forecasted $\lambda_c$, the strategy which we want to test is expected to be in spontaneous arbitrage mode in the next period and the strategy will be executed. When the forecast tells that the strategy would not be in spontaneous arbitrage mode, it will not be exploited and the investor waits until the next execution signals. The weak field is also able to decide the method of portfolio construction. If the constant term is positive, the portfolio which the strategy suggests to build will be constructed. However, if the constant term becomes negative, weights of portfolio will become opposite to those of the portfolio originated from the positive constant term. Simply speaking, the portfolio is not constructed if the speed of adjustment is larger than the critical speed. When it is smaller than the critical value, the weight of the portfolio is $(w_1,w_2,\cdots,w_n)$ if the weak field is positive and the portfolio is $(-w_1,-w_2,\cdots,-w_n)$ if the weak field is negative. This kind of multi-state models is popular in the names of hidden Markov model \cite{Baum:1966} or threshold autoregressive model \cite{Tong:1983} in econometrics and finance. The scheme is repeated in every period over the whole data set.

	To apply the model to real data, the model considered in the continuous time frame needs to be modified to discrete version because all financial data are in the forms of discrete time series. In the discrete form, Eq. (\ref{arb_dyn_ssb}) is changed into
	\begin{eqnarray}
		\label{arb_dyn_ssb_disc}
		r_{i+1}=\big(1-(\lambda-\lambda_c)\big)r_i-\frac{\lambda_c}{r_c^2}r_i^3+\epsilon_{i+1}
	\end{eqnarray}
	and an additional $r_i$ related to the coefficient $1$ in the first term on the r.h.s comes from the time derivative in Eq. (\ref{arb_dyn_ssb}).
	
	The next step is estimation of parameters in Eq. (\ref{arb_dyn_ssb_disc}) with real data. Regression theory gives help on estimation but it is not easy to estimate the parameters with real data because the model is nonlinear and many methods in the regression theory are for linear models. In statistics, these parameters can be estimated by nonlinear regression theory but it is not discussed in this paper. Instead of using nonlinear regression theory directly, we can get some hints from linear regression theory. With consideration on financial meanings and physical dimensions of the parameters, linear regression theory enables us to estimate the model parameters.
	
	There are some issues on the estimation of parameters. The first issue is related to stability of the parameters. When the parameters are fit to real data, if values of the parameters severely fluctuate over time, those abruptly-varying parameters hardly give a steady forecast. One of the best ways to avoid this is taking a moving average (MA) over a certain period. Moving average over the period can make the parameters smoothly-changing parameters. For longer MA windows, the parameter is stable but it would be rather out of date to tell more on the recent situation of the market. If it is short, they can encode more recent information but they tend to vary too abruptly to forecast the future values. To check MA window size dependency, a range of MAs needs to be tested and the results from different MAs should be compared.
		
	Another issue is the method to estimate parameters in the model. Since two or three\footnote{If the weak field is considered, we have three internal parameters.} internal parameters and one external parameter are given in the model, the same number of equations should be prepared. For the simpler case, the coefficient for each term can be considered as one parameter. In this case, two equations need to be set up. However, the values of two parameters found from two equations sometimes diverge when real data are plugged. Since $\lambda$ and $\lambda_c$ are the speeds of adjustment and have same physical and financial meanings, they need to be derived from the same origin. The only difference is that $\lambda_c$ is external. In addition to that, the symmetry breaking needs comparison between two different speeds, $\lambda$ and $\lambda_c$.
	
	One of the possible solutions is that $\lambda$ is derived from the return series of the strategy and $\lambda_c$ comes from the benchmark return as the definition of an external parameter. This interpretation can give two parameters the same physical dimensions and financial meanings. The specification on $\lambda$s is also reasonable in the sense of the efficient market hypothesis. Since the hypothesis tells that it is impossible to systematically outperform the benchmark, it is obvious that we compare the performance of the strategy with that of the benchmark in order to test the hypothesis. In the case of $r^2_c$, the volatility of the strategy or benchmark return can be a good candidate because $r^2_c$ also has the same meaning and dimension to variance. For the constant term $\rho$, the average value of strategy return or benchmark return would be considered. Dividend payment rate is also a good candidate. However, since the most important parameters in the model are $\lambda$ and $\lambda_c$, we focus on the estimation of these two parameters.
	
	The intuitive way to get $\lambda$ and $\lambda_c$ is using a hint from the autoregressive model of order 1 called the AR(1) model. Ignoring the cubic term is also justified by the fact that the returns are much smaller than 1. Starting with the simpler model which does not have the cubic term, multiplying $r_i$ to both sides and taking MA over $k$ periods make the last term zero on the r.h.s. of Eq. (\ref{arb_dyn_ssb_disc}) and give the form of $\lambda$. The one-step ahead forecasted $\lambda$ is
	\begin{eqnarray}
		\hat{\lambda}_{i+1,k}=1-\frac{\langle r_{i}r_{i-1}\rangle_k}{\langle r_{i-1}^2\rangle_k} \nonumber
	\end{eqnarray}
	where $\langle X_{i}\rangle_k=\frac{1}{k}\sum_{j=0}^{k-1}X_{i-j}$. In longer MA windows, we can change $\langle r_{i-1}^2\rangle_k$ in the denominator to $\langle r_{i}^2\rangle_k$ because $\langle r_{i-1}^2\rangle_k$ is close to $\langle r_{i}^2\rangle_k$. In shorter MA windows, the change is meaningful because it is capable of considering more recent informations\footnote{Actually, these two different definitions for $\lambda$ will be tested in next subsections.}. Based on this argument, the final form of forecasted speed of adjustment is given by
	\begin{eqnarray}
		\label{lambda_estimator}
		\hat{\lambda}_{i+1,k}=1-\frac{\langle r_{i}r_{i-1}\rangle_k}{\langle r_{i}^2\rangle_k}.
	\end{eqnarray}
	This $\lambda$ has the same form to the parameter in AR(1) model which is found in
	\begin{eqnarray}
		r_{i+1}&=&\phi r_i +\epsilon_{i+1} \nonumber \\
		\phi&=&\frac{E[x_{i}x_{i-1}]}{E[x_{i}x_{i}]} \nonumber
	\end{eqnarray}	
	where $E[...]$ is the expectation value. 
	
	The estimator (\ref{lambda_estimator}) is intuitively estimated but the hand-weaving argument is available. Since the benchmark return tends to be weakly autocorrelated and the return series by the arbitrage strategy is expected to be strongly positive-autocorrelated, the estimator for the arbitrage strategy is usually smaller than that for the benchmark. In this case, the strategy is in the long-living arbitrage mode. When the estimator for the strategy is larger than that for the benchmark, it is highly probable that return series for the strategy becomes much more weakly autocorrelated than the benchmark return. This tells that the strategy has recently suffered from large downslide and it can be used as the stop signal to strategy execution. Additionally, since the estimator is related to the correlation function which is in the range of -1 and 1, the value of the estimator fluctuates between 0 and 2 and it is well-matched to the positiveness condition on $\lambda$.
	
\subsection{Momentum/Contrarian strategy}
	The momentum strategy is one of the famous trading strategies which use market anomalies. It is well-known that the strategy that buys past winners, short-sells past losers in returns, and then holds the portfolio for some periods in the U.S. market enables to provide positive expected returns in intermediate time frames such as 1--12 months \cite{Jegadeesh:1993p200}. The basic assumption of the strategy is that since price has momentum in its dynamics, it tends to move along the direction it has moved. Based on the assumption, the financial instruments which have shown good performance in the past are highly probable to gain profits in the future. Opposite to winners, it is likely that losers in the past would underperform the benchmark in the monthly time frame. 
		
	Over other trading strategies such as pair trading or merger arbitrage strategies, it is advantageous that the momentum strategy is able to be exploited at any time and in any markets. Pair trading is utilized only when the correlation of two instruments weakens and when investors can find it. Merger arbitrage is able to make benefits if M\&A rumors or news begin to be spread in the market and if there is a price gap between actual and buy prices. When using these strategies, the investors become relatively passive to market conditions and events. However, in the case of the momentum strategy, if they look back at the price history, market participants make use of momentum strategy and the trading frequency is up to their time frames from high frequency trading to long-term investment. In addition to that, unlike merger arbitrage which is possible only in equity markets, momentum strategy can be applied to various asset markets including local equity, global index/equity \cite{Rouwenhosrt:1998,Rouwenhosrt:1999}, currency \cite{Okunev:2003}, commodity \cite{Erb:2006}, future \cite{Asness:2008,Moskowitz:2010}, and bond markets \cite{Asness:2008}.
	
	To exploit momentum strategy, first of all, returns of all equities in the market universe during a certain period called the look-back or ranking period are calculated from closing prices on the first and the last trading dates of the lookback period. Then equities are sorted by their returns in ascending order. Grouped into ten groups, the first group contains the worst performers and the tenth group is for the best performers. Each equity in winner or loser baskets has the same weight in the basket and each basket is also equal in absolute wealth but opposite in position to make the whole portfolio zero cost. The portfolio constructed in zero cost is held during the holding period. The construction of the portfolio occurs on the first day of the holding period and it is liquidated on the last day of the holding period. The transactions happen at the closing prices of each day. The strategy with $J$ period look-back and $K$ period holding period is shortly called $J/K$ strategy.
	
	The expected return by momentum strategy is dependent with lengths of ranking and holding periods. As explained, the strategy with intermediate lengths such as monthly lookback and holding periods generates positive expected return. For longer period strategies, the momentum strategy suffers from reversal such that the winner group loses its price momentum and shows poor performance. Meanwhile, the loser basket shows the opposite behavior such that the basket outperforms and can provide positive return \cite{DeBondt:1985p4232}. According to Lo and MacKinlay \cite{Lo:1990p883}, the short-term momentum strategy in weekly scale also has negative expected return. In both longer and shorter term strategies, it is impossible to make a profit on the momentum portfolio but it does not mean that there is no statistical arbitrage chance nor that the market is efficient. If the portfolio is constructed by reversal momentum i.e. contrarian strategy which buys losers and sells winners, there still exist the chances of profits. The position by contrarian strategy is exactly opposite to that of the momentum strategy.
	
	The reason why the momentum strategy generates positive expected return has attracted the interest of researchers but it is not clearly revealed yet. The sector momentum is considered one of possible explanations \cite{Moskowitz:1999p4294}. A behavioral approach to momentum also can give more explanations such as under-reaction \cite{Hong:1999p4506, Terence:1998p4385} or over-reaction \cite{Daniel:1998p4514} of market participants to news or psychology \cite{Barberisa:1998p307}. It is ambiguous whether the momentum effect comes from either which of them or from a combination of these possible explanations. However, this paper focuses on how to use the strategy based on symmetry breaking rather than what makes markets inefficient.
	
\subsection{Data sets for the strategy}
	Two different market universes are used for analysis to avoid sample selection bias. The first universe is the S\&P 500 index that is the value/float-weighted average of the top 500 companies in market capitalization in NYSE and NASDAQ. It is one of the main indexes such as the Dow Jones Index and Russell 3000 in the U.S. market.  Standard \& Poor's owns and maintains the index with regular rebalancing and component change. Another universe is KOSPI 200 in the South Korean market operated by Korea Exchange (KRX). It is the value-weighted average of 200 companies which represent the main industrial sectors. Unlike the S\&P 500 index, KOSPI 200 contains small-sized companies in market capitalization and considers sector diversification. Its components and weights are maintained regularly and are also irregularly replaced and rebalanced in the case of bankruptcy or upon sector representation issues such as change of core business field or increase/descrease of relative weight in the sector. The significance of each index in the market is much higher than those of other main indexes such as Dow Jones 30 Index or Russell 3000 in the U.S. and the KOSPI index, the value-weighted average of all enlisted equities in the South Korean market, because futures and options on the indexes have enough liquidities to make strong interactions between equity and derivative markets. In the case of the Korean market, the KOSPI 200 index among main indexes is the only index which has the futures and options related to the main indexes. Additionally, many mutual funds target the indexes as their benchmarks and various index-related exchange-traded funds are highly popular in both markets.
	
	The whole time spans considered for two markets are slightly different but have large overlaps. S\&P 500 is covered in the term of Jan. 1st, 1995 and Jun. 30th, 2010, 15.5 years which includes various macro-scale market events such as the Russian/Asian crisis (1997--1998), Dot-com bubble (1995--2000), its collapse (2000--2002), bull market phase (2003-2007), sub-prime mortgage crisis (2007--2009), and the recovery boosted by Quantitative Easing I (2009--2010). In the case of KOSPI 200, the market in the period of Jan. 1st. 2000 to Dec. 31st. 2010, had experienced not only economic coupling to the U.S. market but also local economic turmoils such as the credit bubble and crunch (2002--2003). Given the market and time span, the price history of each stock and whether it was bankrupt or not are stored on a database in order to remove survivor bias and all records for component change are also tracked to keep the number of index components the same. The S\&P 500 data are downloaded from Bloomberg. The whole data of  KOSPI 200 components and their change records are able to be downloaded from KRX. The total number of equities in the database during the covered period is 968 and 411 for S\&P 500, KOSPI 200, respectively\footnote{Unfortunately, S\&P 500 data is not completely free of survivor bias. Histories of 14 equities are not trackable and are left empty in the database. However, it might not give any serious impact on the result because the size of missing data is relatively small compared with the whole dataset.}.
	
\subsection{Results}
\begin{figure}[!h!t!b]
	\subfigure[S\&P 500]{\includegraphics[width=6cm]{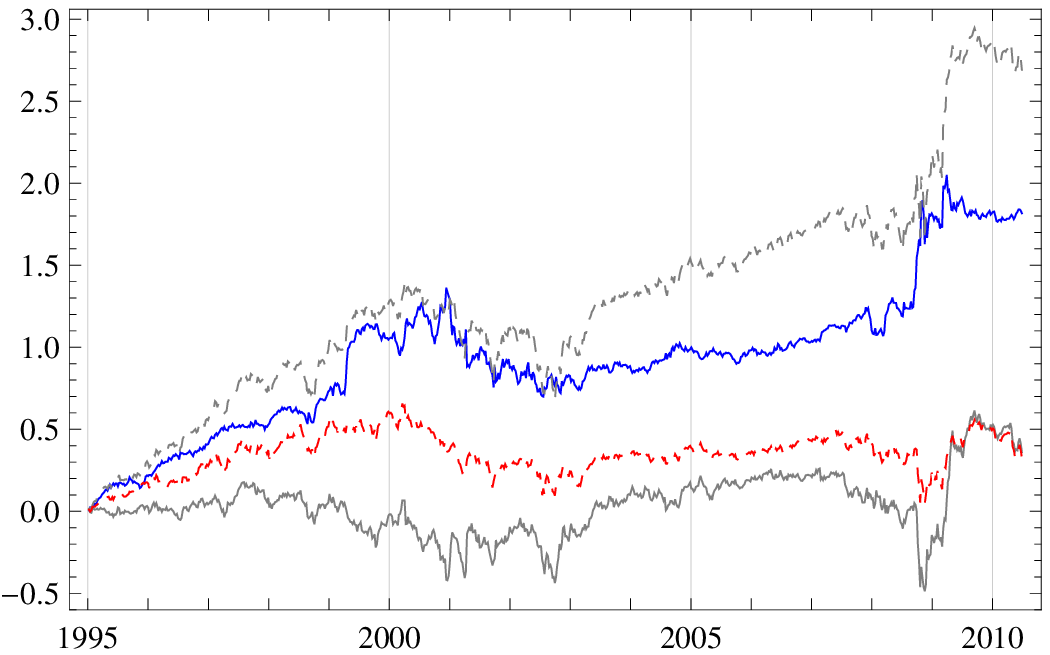}}
	\subfigure[KOSPI 200]{\includegraphics[width=6cm]{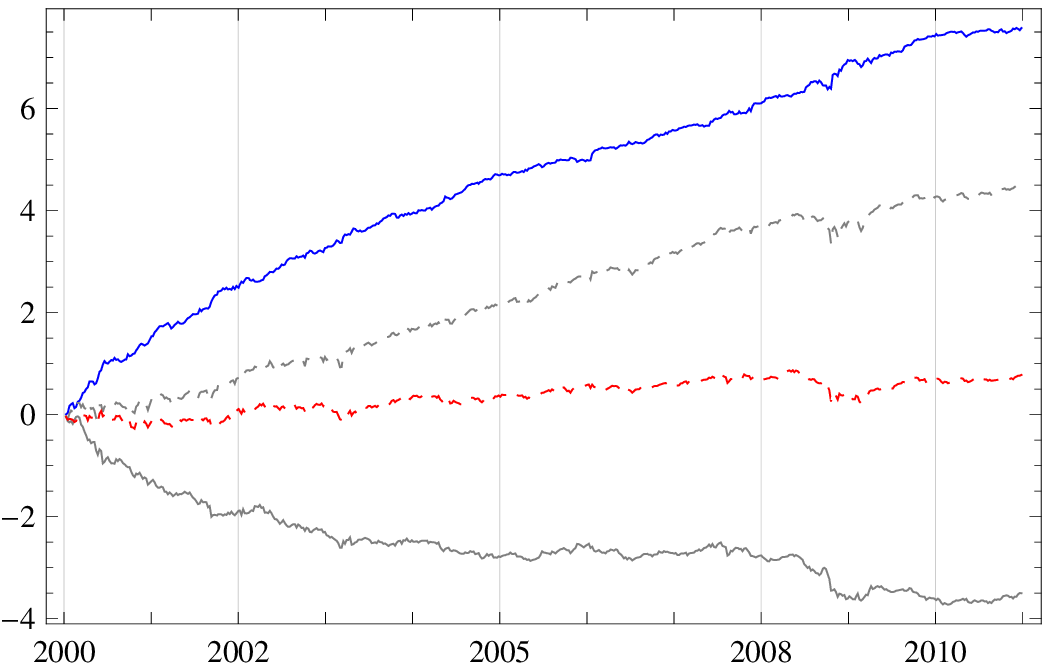}}
	\caption{Cumulative excessive weekly returns in S\&P 500 and KOSPI 200. Return time series by contrarian strategy (blue), by winner (gray), by loser (gray dashed), and by benchmark (red dashed)}
	\label{GRP_Return}		
\end{figure}

	In both markets, $1/1$ weekly strategies are considered and the contrarian portfolios are constructed. The reason for choosing $1/1$ weekly strategies is that they show the best performance in each market among 144 strategies derived from maximum 12-week lookbacks and holdings. Excessive weekly returns of the portfolios are calculated from risk-free rates and proxies for the risk-free rate are from the U.S. Treasury bill with 91 days duration for S\&P 500, CD with 91 days duration for KOSPI 200. Since the weekly momentum portfolio is constructed at the closing price of the first day in the week and is liquidated at the closing price of the last day, the benchmark return is also calculated from the closing prices of the first and the last days in the week. The results for these markets are given in Fig. \ref{GRP_Return}

	There are similarities and differences in two markets. First of all, it is easily seen that $1/1$ strategy shows a reversal that if the winner basket is bought, it is impossible to get a significant positive return but we can achieve positive return from the loser basket. In particular, the contrarian portfolio beats the benchmark over whole periods and this comes from the fact that the loser basket outperforms and the winner basket underperforms the benchmark. Additionally, the contrarian strategy looks more profitable in the Korean market and it can be explained that developed markets have weaker anomalies than emerging markets because the investors in the developed market have utilized the anomalies during longer periods. In the South Korea market, the winner and the loser have more clear directions and magnitudes of the returns are much greater than those of the U.S. market. It is easily seen in Table \ref{TBL_Stat}.
	
\begin{table}[h!t!b!p!]
\tiny
\caption{Statistics for contrarian strategy and benchmark in S\&P 500 and KOSPI 200.}
\begin{tabular}{c c c c c c c c}
\hline
 & & mean & std. & skewness & kurtosis & t-statistics & Sharpe ratio \\ 
\hline
S\&P 500 & Winner & 0.045\% & 3.350\% & 0.419 & 11.181 & 0.379 & 0.013 \\
 & Loser & 0.334\% & 3.973\% & 1.899 & 23.666 & 2.385 & 0.084 \\
 & Contrarian & 0.225\% & 3.097\% & 1.011 & 21.447 & 2.065 & 0.073 \\
 & Benchmark & 0.040\% & 2.368\% & -0.150 & 8.097 & 0.482 & 0.017 \\
\hline
KOSPI 200 & Winner & -0.612\% & 4.189\% & -1.051 & 7.444 & -3.496 & -0.146 \\
 & Loser & 0.796\% & 4.747\% & 0.339 & 9.581 & 4.013 & 0.168 \\
 & Contrarian & 1.325\% & 3.349\% & 1.293 & 9.839 & 9.491 & 0.396 \\
 & Benchmark & 0.136\% & 3.662\% & -0.125 & 7.052 & 0.889 &  0.037\\
\hline
\end{tabular}
\label{TBL_Stat}
\end{table}

	In Table \ref{TBL_Stat}\footnote{The numbers are from excessive weekly return series.}, the numbers from the KOSPI 200 confirm much stronger and clearer contrarian patterns as shown in Fig. \ref{GRP_Return}. The contrarian return in the Korean market is weekly 1.325\% which is much greater than 0.225\% from S\&P 500 contrarian strategy and the t-value of the KOSPI 200 contrarian strategy is 9.491 which is 0.1\% statistically significant but the U.S. strategy has only 2.065, 5\% statistically significant. The null hypothesis is that the expected excessive return is zero. Similar to the contrarian returns, the winner basket and the loser basket have larger absolute returns and t-statistics in KOSPI 200. Both of them are 0.1\% statistically significant but the S\&P 500 loser return only has a 5\% statistically significant t-value and a less significant t-value for another. In both markets, benchmarks have much smaller weekly expected returns than those by the contrarian strategies and t-values are not significant. Standard deviation gives another reason why the portfolio by the momentum/contrarian strategy needs to be constructed. After construction of the contrarian portfolio, the volatility of the portfolio is smaller than the volatility of the winner group and the loser group. In particular, in the South Korea market, the contrarian portfolio has a smaller volatility than the benchmark and has a greater Sharpe ratio than each of the winner and the loser basket has. A larger Sharpe ratio imposes that the strategy is good at minimizing the risk and maximizing the return. Winners, losers, and contrarian portfolios have large kurtosis by fat-tailed distribution.

\begin{figure}[h!t!b!]
	\subfigure[]{\includegraphics[width=6cm]{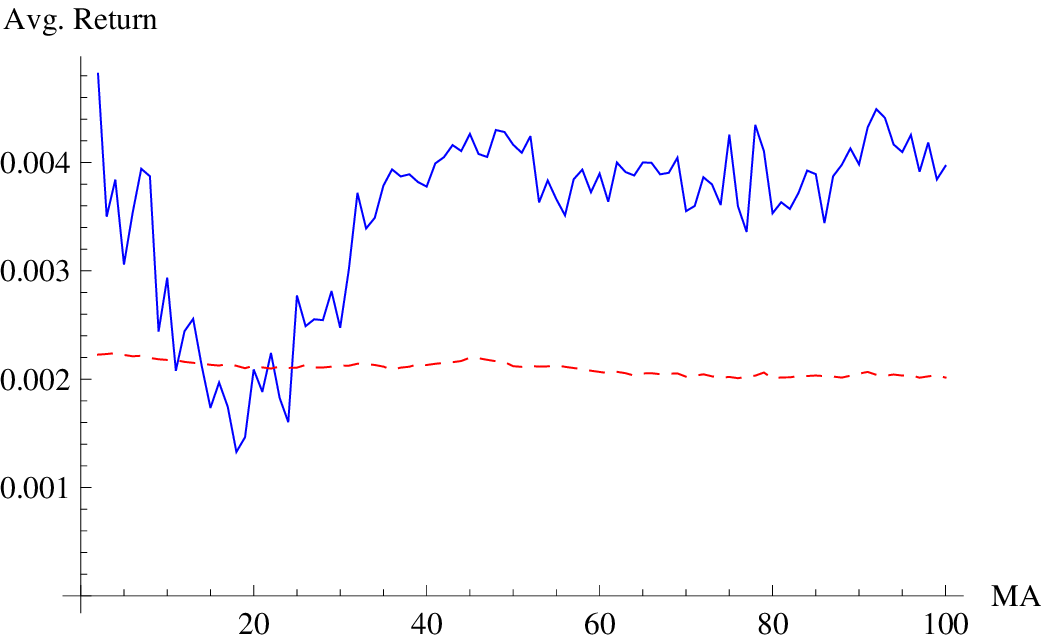}}
	\subfigure[]{\includegraphics[width=6cm]{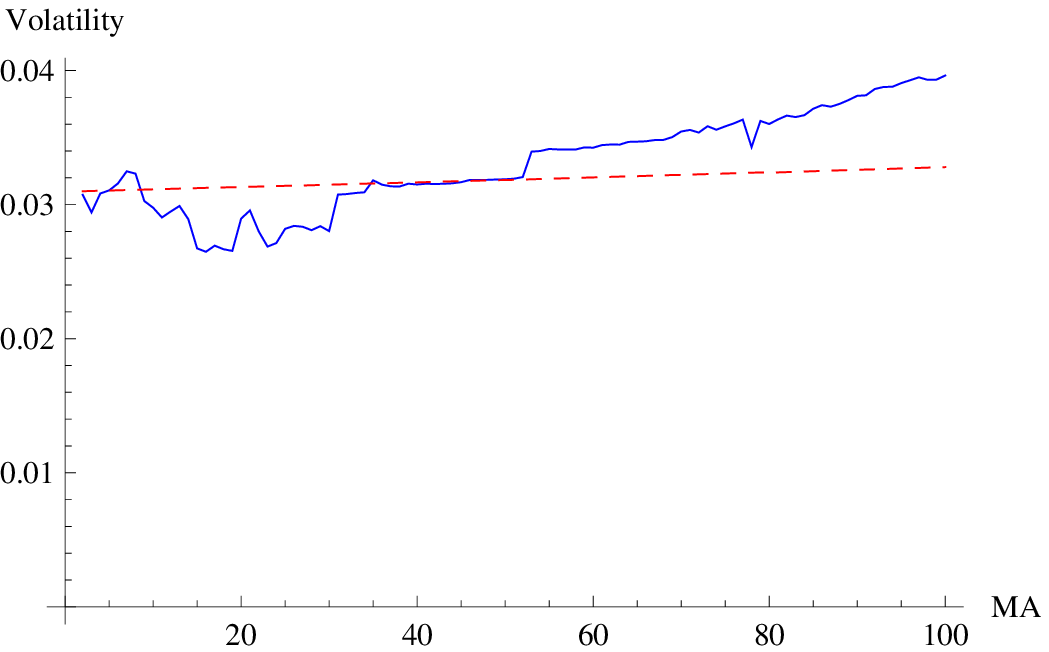}}
	\subfigure[]{\includegraphics[width=6cm]{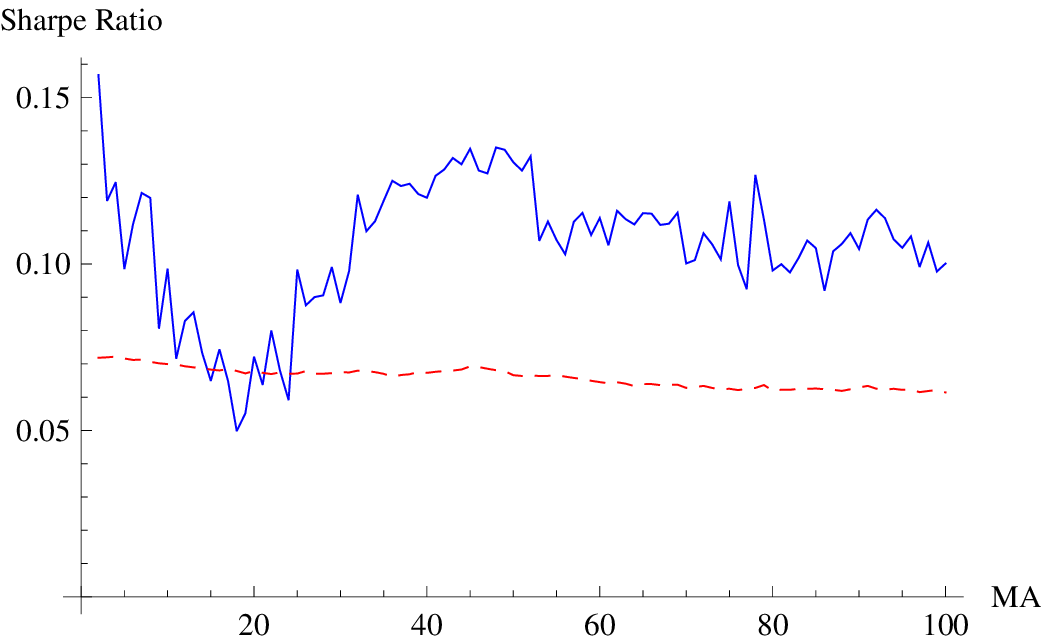}}
	\subfigure[]{\includegraphics[width=6cm]{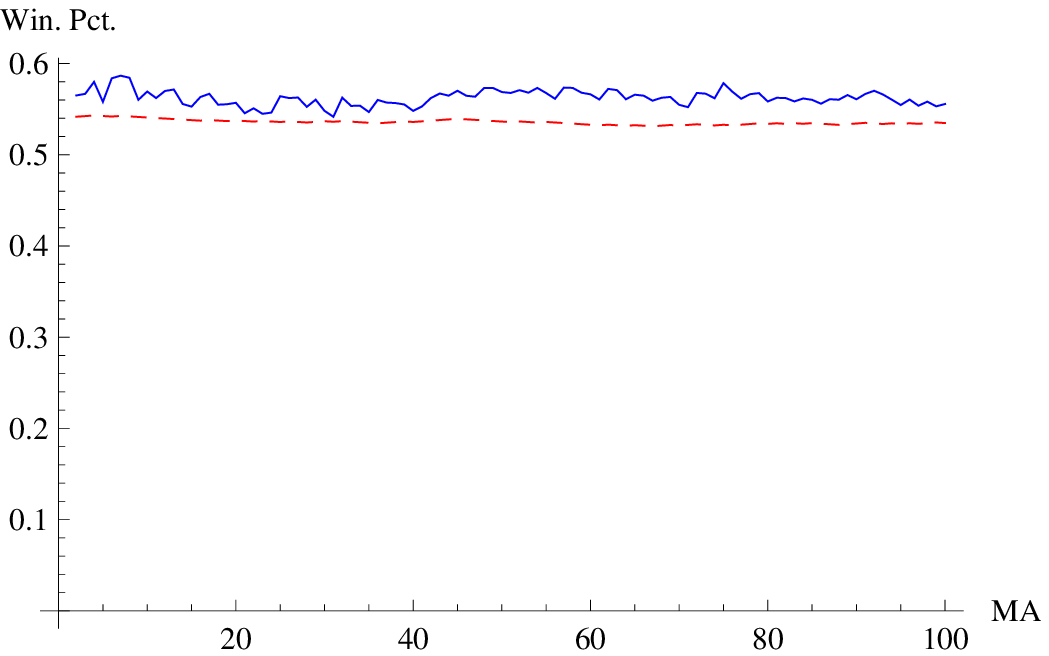}}
	\caption{S\&P 500. SSB-aided momentum strategy (blue) and naive momentum strategy (red dashed). MA window size ranges from 2 to 100.}
	\label{SP500_SSB}		
\end{figure}
\begin{figure}[h]
	\subfigure[]{\includegraphics[width=6cm]{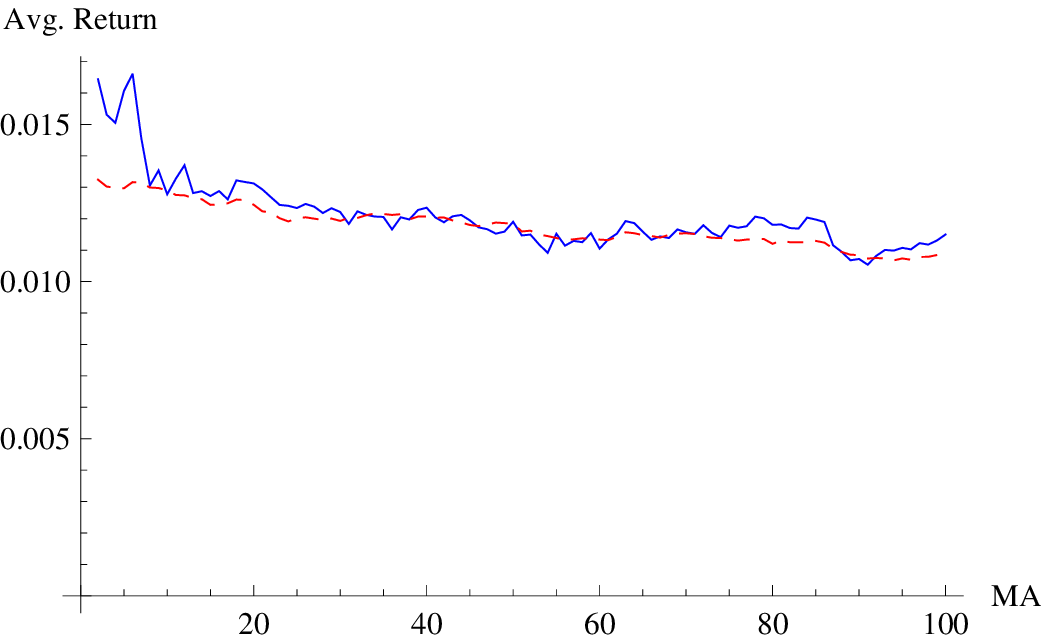}}
	\subfigure[]{\includegraphics[width=6cm]{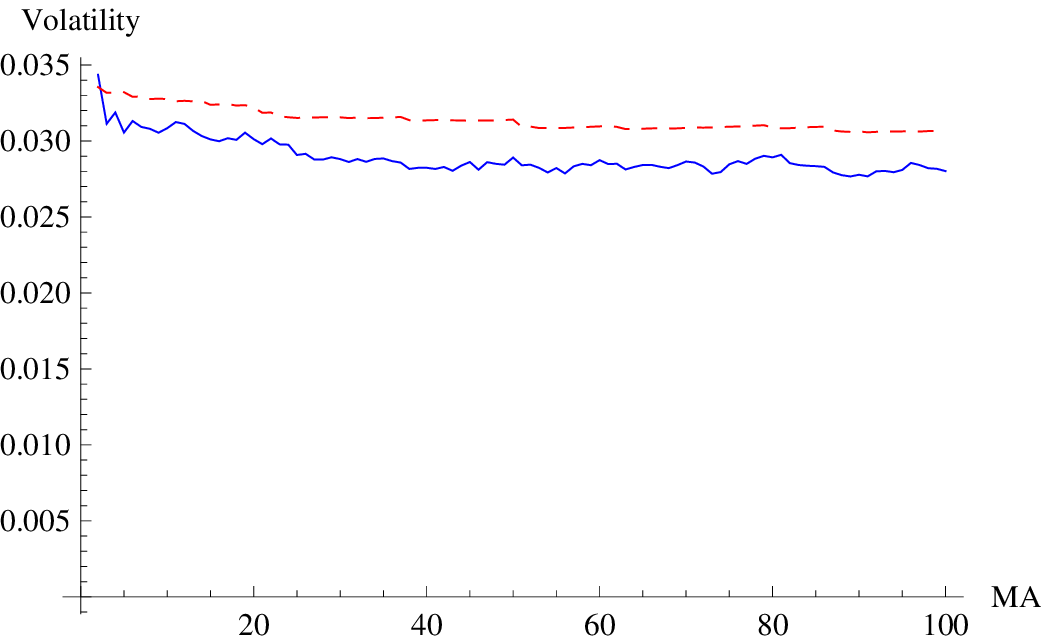}}
	\subfigure[]{\includegraphics[width=6cm]{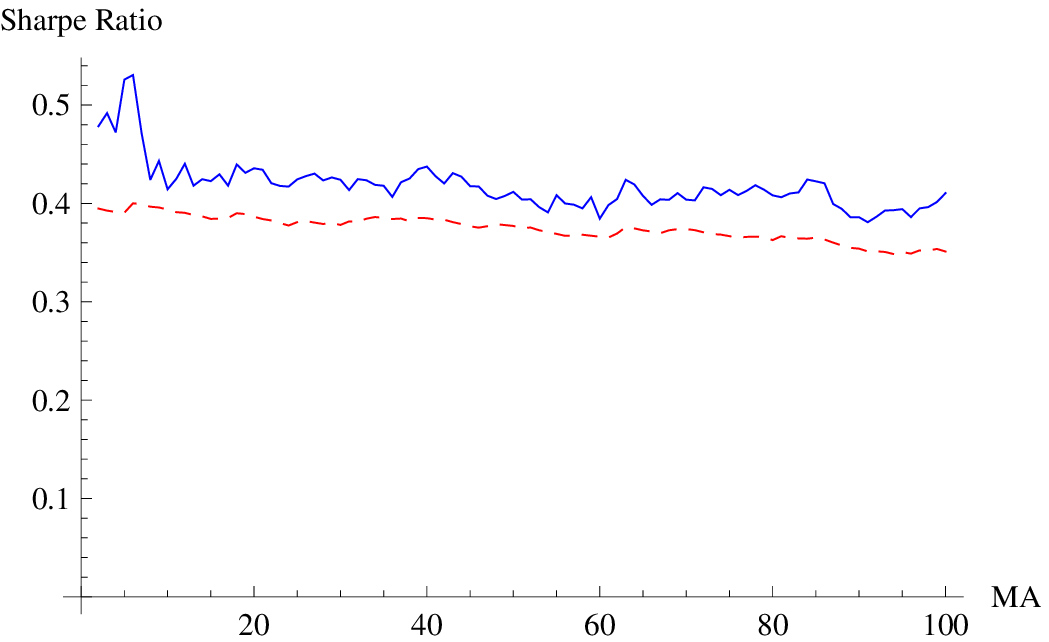}}
	\subfigure[]{\includegraphics[width=6cm]{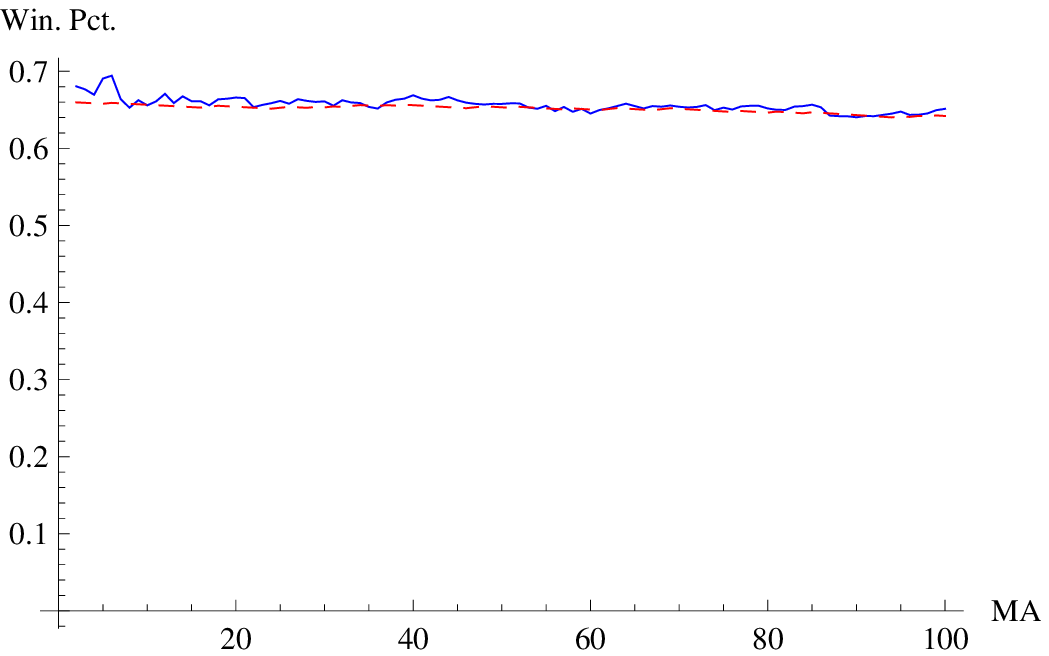}}
	\caption{KOSPI 200. SSB-aided momentum strategy (blue) and naive momentum strategy (red dashed). MA window size ranges from 2 to 100.}
	\label{KOSPI200_SSB}		
\end{figure}

	The results by symmetry breaking with 99 different MA windows are given in Fig. \ref{SP500_SSB} and \ref{KOSPI200_SSB}. The strategies aided by spontaneous symmetry breaking show better performance than the naive momentum strategy in both markets and the results are not particularly dependent on the market where the strategy is used. In the case of return, the strategies with shorter MA windows have improved returns than longer MA windows or naive momentum strategy. As the length of the MA window becomes longer, the return plunges sharply and this plummet is observed in both markets. The Sharpe ratio is also increased with the SSB-aided strategy and it is obvious that the modified strategy is under better risk management. The winning percentage also increases and it is larger for shorter MA windows. 
	
	The application of spontaneous symmetry breaking also has the minor market dependencies. In S\&P 500, average returns and Sharpe ratios increase after a drop around the 20-MA window but the KOSPI 200 momentum does not recover its average return level and remains stagnated around returns by the naive strategies. In the case of volatility, it is helpful to reduce volatility with SSB in KOSPI 200 but is not useful in S\&P 500. 

	The constant term in spontaneous symmetry breaking is also considered. As described before, average return over the MA window or return in previous term of the raw strategy are used as the forecasted constant term. If the constant is positive, the contrarian portfolio is constructed and if the constant is negative, the momentum strategy is used. However, the strategy including the constant term does not provide better results than the strategy without the constant. The same approach is applied to mean return or return in previous terms of the benchmark but it is not possible to find the better strategy. With these facts, it is guessed that the constant term is zero or the constant term is always positive if it exists and if these returns are the only possible candidates for the constant. The positiveness of the constant can be guaranteed by the fact that the arbitrage portfolio is constructed to get a positive expected return.
	
	With other estimators for speed of adjustment, it is found that the SSB-guided strategies provide similar results although the results are not given in the paper. In both markets, the patterns of results are similar to the results depicted in Fig. \ref{SP500_SSB} and \ref{KOSPI200_SSB}. Specifically speaking, the estimator with $\langle r_{i-1}^2\rangle_k$ in the denominator gives similar patterns in KOSPI 200 but the performance is slightly poorer than the result in Fig. \ref{KOSPI200_SSB}. In the U.S. market, similar patterns in longer MA windows are found but the results with shorter MA windows are worse than the result given in the paper. This is well-matched to the assumption that $\langle r_{i-1}^2\rangle_k$ is almost identical to $\langle r_{i}^2\rangle_k$ in longer MA windows. When the estimator uses the covariance of $r_i$ and $r_{i-1}$, similar results are found but the performance becomes much poorer, especially in shorter window length.
	
	Although the whole time period is same for each of the MA windows, longer MA strategies have fewer data points when the performance is calculated in backtest. This difference in number of data points comes from the assumption that even though we work with historical data already known, we pretend to be unaware of the future after the moment at which the forecast is made in backtest. In the simulation with each MA, the first few data whose length is the same as the size of the MA window are used for forecast and are ignored in the calculation of performance. However, the difference does not make any serious difference in the patterns of performance. When the tests to calculate the performance are repeated over the same sample period for all MA windows, notable differences are not observed and the results are similar to Fig. \ref{SP500_SSB} and \ref{KOSPI200_SSB}.

\section{Conclusion}
\label{Conclu}
	The cubic order term and parity symmetry on return introduce the concept of spontaneous symmetry breaking to arbitrage dynamics. In the asymptotic time region, the dynamics has symmetry breaking modes triggered by the control parameter. It can provide the long-living arbitrage modes including the short-living mode in the dynamics. Spontaneous symmetry breaking generated by the control parameter $\lambda$ imposes phase transition between the arbitrage phase and no-arbitrage phase. Contrasting to the short-living mode which is expected in the frame of the efficient market hypothesis, the long-living modes are totally new and exotic. The existence of a spontaneous arbitrage mode explains why the arbitrage return survives longer than expected and why the trading strategies based on market anomalies can make long term profits. With the existence of the weak field, it is possible to consider the transition between two long-living arbitrage modes, $\pm r_v$ in the asymptotic region.
	
	Based on spontaneous symmetry breaking of arbitrage, the control parameter enables to decide execution of the trading strategy. If $\lambda$ for the strategy is smaller than $\lambda_c$ for the benchmark, the strategy will be executed in next period. If the speed of adjustment for the strategy is greater than that of the benchmark, nothing will be invested. Since it is difficult to estimate the parameter in the nonlinear model, the AR(1) model gives an insight for estimation. The estimated $\lambda$ based on the AR(1) model has the theoretical ground that the speed of estimation is derived from the autocorrelation function. It is also reasonable in the sense of testing the efficient market hypothesis because it is capable of comparing the strategy with the benchmark. The simplest but most meaningful estimator for the control parameter is applied to momentum strategy in the U.S and South Korean stock markets. The SSB-aided momentum strategy outperforms and has lower risk than the naive momentum strategy has. Since the strategy applied to two different markets shows similar patterns, the results are not achieved by data snooping. It is also not by estimator bias because three different estimators for speed of adjustment are tested and provide similar results with some minor differences.
		
	The future study will be stretched into a few directions. First of all, parameter estimation needs to be more precise and statistically meaningful. In this paper, the estimator for the control parameter $\lambda$ is from the AR(1) model and $\lambda$ for benchmark serves as the critical value $\lambda_c$ although the cubic term exists. Although it provides better performance and lower risk, estimation of the parameters is from the reasonable intuition not from regression theory. For the more precise model, they need to be estimated from nonlinear regression theory. In particular, a statistical test on estimation should be done. In the case of $\lambda_c$, it can be estimated with the help of other researches on market phase such as Wyarta and Bouchaud's work \cite{Wyarta:2007}. Other parameters, $r_c$ or $\rho$, also help to find the better performance strategy if they are statistically well-estimated. The second direction is considering the stochastic term in arbitrage dynamics. In the paper, only the deterministic part is considered and the stochastic term is out of interest in finding the exact solutions. If the spontaneous symmetry breaking modes are found not as the asymptotic solutions but as the exact solutions of the stochastic differential equation, they would extend our understanding on arbitrage dynamics. In addition to that, specification of relaxation time can be found from the correlation function of the stochastic solutions. Finally, it would be interesting if validity of the arbitrage modeling with spontaneous symmetry breaking is tested over other arbitrage strategies. Since only the momentum/contrarian strategy is the main concern in the paper, tests on other trading strategies including high frequency trading look very interesting. Additionally, a cross-check with momentum strategies for different markets and frequencies would be helpful to check the effectiveness and usefulness of spontaneous symmetry breaking concepts in arbitrage modeling.

\section*{Acknowledgements}
	It is our pleasure to thank Matthew Atwood, Jonghyoun Eun, Robert J. Frey, Wonseok Kang, Andrew Mullhaupt, and Svetlozar Rachev for useful discussions. We are especially indebted to Sungsoo Choi for helpful discussions from the early stage of this work. We are grateful to Didier Sornette for providing valuable advice in the revision of the first draft. We express thanks to Sungsoo Choi, Jonghyoun Eun, and Wonseok Kang for cooperation on the construction of the financial database on the Korean stock market. We are thankful to Xiaoping Zhou for collecting parts of price histories on S\&P 500 components.



\begin{thebibliography}{99}
\bibitem{Bachelier:1900p4113}
	Bachelier, L., Theorie de la spéculation, Annales Scientifiques de l’École Normale Supérieure 3 (1900) 21-86
\bibitem{Cootner:1964p4779}
	Cootner, P., The random character of stock market prices, MIT Press, 1964
\bibitem{Black:1973p5045}
	Black, F. and Scholes M. The pricing of options and corporate liabilities, Journal of Political Economy 8 (1973) 637-654
\bibitem{Merton:1973p5006}
	Merton, R. C., Theory of rational option pricing, Bell Journal of Economics and Management Science (The RAND Corporation) 4 (1973) 141-183.
\bibitem{Vasicek:1977p5439}
	Vasicek, O., An equilibrium characterization of the term structure, Journal of Financial Economics 5 (1977) 177-188
\bibitem{Cox:1985p5352}
	Cox, J.C., Ingersoll, J.E. and Ross, S. A., A theory of the term structure of interest rates, Econometrica 53 (1985) 385-407
\bibitem{White:1990p5282}
	Hull, J. and White, A., Pricing interest-rate derivative securities, The Review of Financial Studies 3 (1990) 573-592
\bibitem{Fama:1965p4897}
	Fama, E., The behavior of stock market prices, Journal of Business 38 (1965) 34-105 
\bibitem{Samuelson:1965p4904}
	Samuelson, P., Proof that properly anticipated prices fluctuate randomly, Industrial Management Review 6 (1965) 41-49 
\bibitem{Singal:2006p5475}
	Singal, V., Beyond the random walk: A guide to stock market anomalies and low risk investing, Oxford University Press, 2003
\bibitem{Lo:2001p5538}
	Lo, A. and MacKinlay, A., A non-random walk down wall street, Princeton University Press, 2001
\bibitem{Shleifer:2000}
	Shleifer, A., Inefficient markets : An introduction to behavioral finance, Oxford University Press, 2000
\bibitem{Kahneman:2000}
	Kahneman, D. and Tversky, A., Choices, values and frames, Cambridge University Press, 2000
\bibitem{Kahneman:1982}
	Kahneman, D., Slovic, P., and Tversky, A. Judgment under uncertainty: Heuristics and biases, Cambridge University Press, 1982 
\bibitem{Conlisk:1996p1448}
	Conlisk, J., Bounded rationality and market fluctuation, Journal of Economic Behavior and Organization 29 (1996) 233-250	
\bibitem{Jegadeesh:1993p200}
	Jegadeesh, N. and Titman, S., Returns to buying winners and selling losers: Implications for stock market efficiency, Journal of Finance 48 (1993) 65-91	
\bibitem{Mantegna}
	Mangtegna, R. N. and Stanley, H. E., Introduction to econophysics: Correlations and complexity in finance, Cambridge University Press, 2007
\bibitem{Roehner}
	Roehner, B. M., Patterns of Speculation: A study in observational econophysics, Cambridge University Press, 2005
\bibitem{Sornette}
	Sornette, D., Why Stock Markets Crash: Critical events in complex financial systems, Princeton University Press, 2004
\bibitem{Sornette:2001p32}
	Sornette, D. and Malevergne, Y., From rational bubbles to crashes, Physica A 299 (2001) 40-59 (2001)
\bibitem{Malevergne:2001}
	 Malevergne, Y. and Sornette, D., Multi-dimensional rational bubbles and fat tails, Quantitative Finance 1 (2001) 533-541
\bibitem{Sornette:2000p4733}
	Sornette, D., Stock market speculation : Spontaneous symmetry breaking of economic valuation, Physica A  284  (2000) 355-375
\bibitem{Wyarta:2007}
	Wyarta, M. and Bouchaud, J. P., Self-referential behaviour, overreaction and conventions in financial markets, Journal of Economic Behavior and Organization 63 (2007) 1-24
\bibitem{Girsanov}
	Girsanov, I. V., On transforming a certain class of stochastic processes by absolutely continuous substitution of measures, Theory Prob. Appl. 5 (1960) 285-301
\bibitem{Ilinski:2001p4046}
	Ilinski, K., Physics of finance: Gauge modelling in non-equilibrium pricing, Wiley, 2001
\bibitem{Cochrane:2011}
	Cochrane, John H., Presidential address: Discount rates, Journal of Finance 66 (2011) 1047-1108
\bibitem{Amihud}
	Amihud, Y. and Mendelson, H., Trading mechanisms and stock returns: An empirical investigation, Journal of Finance 42 (1987) 533-553 
\bibitem{Damodaran}
	Damodaran, A., A simple measure of price adjustment coefficients, Journal of Finance 48 (1993) 387-400 
\bibitem{Theobald:1998}
	Theobald, M., and Yallup, P., Measuring cash-futures temporal effects in the UK using partial adjustment factors, Journal of Banking and Finance 22 (1998) 221-243 
\bibitem{Theobald:1999}
	Theobald, M., and Yallup, P., Determining security speed of adjustment coefficient, Journal of Financial Markets 7 (2004) 75-96
\bibitem{Bouchaud:1998p1217}
	Bouchaud, J. P. and Cont R., A Langevin approach to stock market fluctuations and crashes, European Physical Journal B  6 (1998) 543-550
\bibitem{Hong:1999p4506}
	Hong, H. and Stein, J. C., A unified theory of underreaction, momentum trading and overreaction in asset markets, Journal of Finance 54 (1999) 2143-2184
\bibitem{Terence:1998p4385}
	Terence, H., Hong, H., Lim, T., and Stein, J., Bad news travels slowly: Size, analyst coverage, and the profitability of momentum strategies, Journal of Finance 55 (2000) pp. 265-295
\bibitem{Baum:1966}
	Baum, L. E. and Petrie, T. Statistical Inference for Probabilistic Functions of Finite State Markov Chains. The Annals of Mathematical Statistics 37 (1966): 1554-1563
\bibitem{Tong:1983}
	Tong, H., Threshold Models in Nonlinear Time Series Analysis, Springer-Verlag, New York, 1983
\bibitem{Rouwenhosrt:1998}
	Rouwenhorst, K. G., International momentum strategies, Journal of Finance 53 (1998) 267-284
\bibitem{Rouwenhosrt:1999}
	Rouwenhorst, K. G., Local return factors and turnover in emerging stock markets, Journal of Finance 54 (1999) 1439-1464
\bibitem{Okunev:2003}
	Okunev, John, and Derek White, Do momentum-based strategies still work in foreign currency markets?, Journal of Financial and Quantitative Analysis 38 (2003) 425-447
\bibitem{Erb:2006}
	Erb, Claude B., and Campbell R. Harvey, The strategic and tactical value of commodity futures, Financial Analysts Journal 62 (2006) 69-97
\bibitem{Asness:2008}
	Asness, C. S., Moskowitz, T., and Pedersen, L. H., Value and momentum everywhere, University of Chicago working paper, 2008
\bibitem{Moskowitz:2010}
	Moskowitz, T. J., Ooi, Y. H., and Pedersen, L. H., Time series momentum, University of Chicago Working Paper, 2010
\bibitem{DeBondt:1985p4232}
	De Bondt, W. F. M. and Thaler, R., Does the stock market overreact?, Journal of Finance  40 (1985) 793-805
\bibitem{Lo:1990p883}
	Lo, A. and MacKinlay, A., When are contrarian profits due to stock market overreaction?, Review of Financial Studies 3 (1990) 175-205 
\bibitem{Moskowitz:1999p4294}
	Moskowitz, T. J. and Grinblatt M., Do industries explain momentum?, Journal of Finance 54 (1999) 1249-1290
\bibitem{Daniel:1998p4514}
	 Daniel, K., Hirshleifer, D., Subrahmanyam, A., Investor psychology and security market under- and over-reactions. Journal of Finance 53 (1998) 1839-1886
\bibitem{Barberisa:1998p307}
	Barberis, N., Shleifer, A., Vishny, R. A model of investor sentiment. Journal of Financial Economics 49 (1998) 307-343


\end{thebibliography}
\end{document}